\input harvmac
\def\half{{\textstyle {1 \over 2}}}

\overfullrule=0pt 
\parindent 25pt
\tolerance=10000
\sequentialequations
%\draftmode

%%%%%%%References%%%%%%%
\def\lr{\lref}

\lr\vafaa{L.~J.~Dixon, V.~Kaplunovsky and C.~Vafa,
``On Four-Dimensional Gauge Theories From Type II Superstrings,''
Nucl.\ Phys.\  {\bf B294} (1987) 43.
%%CITATION = NUPHA,B294,43;%%
}
\lr\dolan{R.~Bluhm, L.~Dolan and P.~Goddard,
``A New Method Of Incorporating Symmetry Into Superstring Theory,''
Nucl.\ Phys.\  {\bf B289} (1987) 364.
%%CITATION = NUPHA,B289,364;%%
}

\lr\ferraraa{S.~Ferrara and C.~Kounnas,
``Extended Supersymmetry In Four-Dimensional Type Ii Strings,''
Nucl.\ Phys.\  {\bf B328} (1989) 406.
%%CITATION = NUPHA,B328,406;%%
}

\lr\bianchia{M.~Bianchi, E.~Gava, F.~Morales and K.~S.~Narain,
``D-strings in unconventional type I vacuum configurations,''
Nucl.\ Phys.\  {\bf B547} (1999) 96
[hep-th/9811013].
%%CITATION = HEP-TH 9811013;%%
}

\lr\antona{I.~Antoniadis, C.~P.~Bachas and C.~Kounnas,
``Four-Dimensional Superstrings,''
Nucl.\ Phys.\  {\bf B289} (1987) 87.
%%CITATION = NUPHA,B289,87;%%
}

\lr\gimona{E.~G.~Gimon and J.~Polchinski,
``Consistency Conditions for Orientifolds and D-Manifolds,''
Phys.\ Rev.\  {\bf D54} (1996) 1667
[hep-th/9601038].
%%CITATION = HEP-TH 9601038;%%
}

\lr\parka{A.~Dabholkar and J.~Park,
``Strings on Orientifolds,''
Nucl.\ Phys.\  {\bf B477} (1996) 701
[hep-th/9604178].
%%CITATION = HEP-TH 9604178;%%
}

\lr\banksa{T.~Banks and L.~J.~Dixon,
``Constraints On String Vacua With Space-Time Supersymmetry,''
Nucl.\ Phys.\  {\bf B307} (1988) 93.
%%CITATION = NUPHA,B307,93;%%
}

\lr\pola{J.~Polchinski,
``Dirichlet-Branes and Ramond-Ramond Charges,''
Phys.\ Rev.\ Lett.\  {\bf 75} (1995) 4724
[hep-th/9510017].
%%CITATION = HEP-TH 9510017;%%
}

\lr\cardy{J.~L.~Cardy,
``Boundary Conditions, Fusion Rules And The Verlinde Formula,''
Nucl.\ Phys.\  {\bf B324} (1989) 581.
%%CITATION = NUPHA,B324,581;%%
}

\lr\devech{P.~Di Vecchia and A.~Liccardo,
``D branes in string theory. I,''
hep-th/9912161.
%%CITATION = HEP-TH 9912161;%%
}

\lr\tyea{H.~Kawai, D.~C.~Lewellen and S.~H.~Tye,
``Classification Of Closed Fermionic String Models,''
Phys.\ Rev.\  {\bf D34} (1986) 3794.
%%CITATION = PHRVA,D34,3794;%%
}

\lr\sena{A.~Sen,
``Non-BPS states and branes in string theory,''
hep-th/9904207.
%%CITATION = HEP-TH 9904207;%%
}

\lr\gaberdiela{M.~R.~Gaberdiel,
``Lectures on non-BPS Dirichlet branes,''
hep-th/0005029.
%%CITATION = HEP-TH 0005029;%%
}

\lr\schwarza{J.~H.~Schwarz,
``TASI lectures on non-BPS D-brane systems,''
hep-th/9908144.
%%CITATION = HEP-TH 9908144;%%
}

\lr\dixona{L.~Dixon, J.~A.~Harvey, C.~Vafa and E.~Witten,
``Strings On Orbifolds,''
Nucl.\ Phys.\  {\bf B261} (1985) 678.
%%CITATION = NUPHA,B261,678;%%
}

\lr\vafab{C.~Vafa,
``Modular Invariance And Discrete Torsion On Orbifolds,''
Nucl.\ Phys.\  {\bf B273} (1986) 592.
%%CITATION = NUPHA,B273,592;%%
}

\lr\callana{
C.~G.~Callan, C.~Lovelace, C.~R.~Nappi and S.~A.~Yost,
``Adding Holes And Crosscaps To The Superstring,''
Nucl.\ Phys.\  {\bf B293} (1987) 83.
%%CITATION = NUPHA,B293,83;%%
}

\lr\greena{M.~B.~Green,
``Pointlike states for type 2b superstrings,''
Phys.\ Lett.\  {\bf B329} (1994) 435
[hep-th/9403040].
%%CITATION = HEP-TH 9403040;%%
}

\lr\greenb{M.~B.~Green and M.~Gutperle,
``Light-cone supersymmetry and D-branes,''
Nucl.\ Phys.\  {\bf B476} (1996) 484
[hep-th/9604091].
%%CITATION = HEP-TH 9604091;%%
}

\lr\greend{M.~B.~Green and M.~Gutperle,
``Symmetry Breaking at enhanced Symmetry Points,''
Nucl.\ Phys.\  {\bf B460} (1996) 77
[hep-th/9509171].
%%CITATION = HEP-TH 9509171;%%
}

\lr\senc{A.~Sen,
``SO(32) spinors of type I and other solitons on brane-antibrane pair,''
JHEP {\bf 9809} (1998) 023
[hep-th/9808141].
%%CITATION = HEP-TH 9808141;%%
}

\lr\polc{E.~G.~Gimon and J.~Polchinski,
``Consistency Conditions for Orientifolds and D-Manifolds,''
Phys.\ Rev.\  {\bf D54} (1996) 1667
[hep-th/9601038].
%%CITATION = HEP-TH 9601038;%%
}

\lr\send{A.~Sen,
``Duality and Orbifolds,''
Nucl.\ Phys.\  {\bf B474} (1996) 361
[hep-th/9604070].
%%CITATION = HEP-TH 9604070;%%
}

\lr\wittend{
C.~Vafa and E.~Witten,
``Dual string pairs with N = 1 and N = 2 supersymmetry in four  dimensions,''
Nucl.\ Phys.\ Proc.\ Suppl.\  {\bf 46} (1996) 225
[hep-th/9507050].
%%CITATION = HEP-TH 9507050;%%
}

\lr\bianchia{M.~Bianchi,
``A note on toroidal compactifications of the type I superstring and  other superstring vacuum configurations with 16 supercharges,''
Nucl.\ Phys.\  {\bf B528} (1998) 73
[hep-th/9711201].
%%CITATION = HEP-TH 9711201;%%
}

\lr\bianchib{M.~Bianchi, E.~Gava, F.~Morales and K.~S.~Narain,
``D-strings in unconventional type I vacuum configurations,''
Nucl.\ Phys.\  {\bf B547} (1999) 96
[hep-th/9811013].
%%CITATION = HEP-TH 9811013;%%
}

\lr\wittenf{E.~Witten,
``Toroidal compactification without vector structure,''
JHEP {\bf 9802} (1998) 006
[hep-th/9712028].
%%CITATION = HEP-TH 9712028;%%
}

\lr\gregoria{A.~Gregori, E.~Kiritsis, C.~Kounnas, N.~A.~Obers, P.~M.~Petropoulos and B.~Pioline,
``$R^2$ corrections and non-perturbative dualities of N = 4 string ground  states,''
Nucl.\ Phys.\  {\bf B510} (1998) 423
[hep-th/9708062].
%%CITATION = HEP-TH 9708062;%%
}

\lr\bagrsch{C.~P.~Bachas, M.~B.~Green and A.~Schwimmer,
``(8,0) quantum mechanics and symmetry enhancement in type I'  superstrings,''
JHEP {\bf 9801} (1998) 006
[hep-th/9712086].
%%CITATION = HEP-TH 9712086;%%
}

\lr\brunnera{I.~Brunner, A.~Rajaraman and M.~Rozali,
``D-branes on asymmetric orbifolds,''
Nucl.\ Phys.\  {\bf B558} (1999) 205
[hep-th/9905024].
%%CITATION = HEP-TH 9905024;%%
}

\lr\bianchic{M.~Bianchi, J.~F.~Morales and G.~Pradisi,
``Discrete torsion in non-geometric orbifolds and their open-string  descendants,''
Nucl.\ Phys.\  {\bf B573} (2000) 314
[hep-th/9910228].
%%CITATION = HEP-TH 9910228;%%
}

\lr\carloa{C.~Angelantonj, R.~Blumenhagen and M.~R.~Gaberdiel,
``Asymmetric orientifolds, brane supersymmetry breaking and non-BPS  branes,''
hep-th/0006033.
%%CITATION = HEP-TH 0006033;%%
}

\lr\blumenhagenb{R.~Blumenhagen, L.~Gorlich, B.~Kors and D.~Lust,
``Asymmetric orbifolds, noncommutative geometry and type I string vacua,''
hep-th/0003024.
%%CITATION = HEP-TH 0003024;%%
}

\lr\wittenf{
J.~Polchinski and E.~Witten,
``Evidence for Heterotic - Type I String Duality,''
Nucl.\ Phys.\  {\bf B460} (1996) 525
[hep-th/9510169].
%%CITATION = HEP-TH 9510169;%%
}

\lr\witteng{
E.~Witten,
``Toroidal compactification without vector structure,''
JHEP {\bf 9802} (1998) 006
[hep-th/9712028].
%%CITATION = HEP-TH 9712028;%%
}

\lr\bianchig{M.~Bianchi,
``A note on toroidal compactifications of the type I superstring and  other superstring vacuum configurations with 16 supercharges,''
Nucl.\ Phys.\  {\bf B528} (1998) 73
[hep-th/9711201].
%%CITATION = HEP-TH 9711201;%%
}

\lr\bianchih{M.~Bianchi, G.~Pradisi and A.~Sagnotti,
``Toroidal compactification and symmetry breaking in open string theories,''
Nucl.\ Phys.\  {\bf B376} (1992) 365.
%%CITATION = NUPHA,B376,365;%%
}

\lr\polcai{J.~Polchinski and Y.~Cai,
``Consistency Of Open Superstring Theories,''
Nucl.\ Phys.\  {\bf B296} (1988) 91.
%%CITATION = NUPHA,B296,91;%%
}

\lr\tyec{H.~Kawai, D.~C.~Lewellen and S.~H.~Tye,
``Construction Of Fermionic String Models In Four-Dimensions,''
Nucl.\ Phys.\  {\bf B288} (1987) 1.
%%CITATION = NUPHA,B288,1;%%
}

\lr\sagnottik{I.~Antoniadis, E.~Dudas and A.~Sagnotti,
``Supersymmetry breaking, open strings and M-theory,''
Nucl.\ Phys.\  {\bf B544} (1999) 469
[hep-th/9807011].
%%CITATION = HEP-TH 9807011;%%
}

%%%%%%%%%%%%%%%%%%%%%

\lr\bgh{O. Bergman, E. Gimon and  P. Horava, {\it
Brane Transfer Operations and T-Duality of Non-BPS States}, 
JHEP {\bf 9904} (1999) 010, hep-th/9902160 .}

\lr\sagnotti{M. Bianchi, G. Pradisi and A. Sagnotti, {\it
Toroidal Compactification and Symmetry Breaking in Open String
Theories}, Nucl. Phys. {\bf 376} (1992) 365.}

\lr\wittena{
E. Witten, {\it Torroidal Compactification without Vector Structure},
JHEP {\bf 9802} (1998) 006, hep-th/9712028 . }

\lr\lerdar{
A.~Lerda and R.~Russo,
``Stable non-BPS states in string theory: A pedagogical review,''
Int.\ J.\ Mod.\ Phys.\  {\bf A15} (2000) 771
[hep-th/9905006].
%%CITATION = HEP-TH 9905006;%%}
}

%%%%%%%%%%%%%%%%%%%%%%%%%%%%%%%%%%%%%%%%%%%%%%%%%%%%%%%%%%%%%%%%%%%
%%%%%%%%% title and abstract
%%%%%%%%%%%%%%%%%%%%%%%%%%%%%%%%%%%%%%%%%%%%%%%%%%%%%%%%%%%%%%%%%%%
\noblackbox
\baselineskip 20pt plus 2pt minus 2pt
\Title{\vbox{\baselineskip12pt 
\hbox{hep-th/0007126}
\hbox{HUTP-00/A026}
  }}
{\vbox{\centerline{Non-BPS D-branes and enhanced symmetry}\medskip
 \centerline{in an
 asymmetric orbifold}}}  

\centerline{Michael Gutperle}
\smallskip
\centerline{\it Jefferson Laboratory of Physics}
\centerline{\it Harvard University, Cambridge, MA 02138, USA}
\bigskip
\centerline{\tt gutperle@riemann.harvard.edu}
\bigskip

%% abstract
\medskip
\centerline{{\bf Abstract}}
\noindent In this paper   properties of D branes in a nine
dimensional asymmetric orbifold are discussed, using a
$(-1)^{F_L}\sigma_{1/2}$ projection, 
where $F_L$ is the leftmoving space-time fermion number and $\sigma_{1/2}$ is
a freely acting shift of order two. There are two types of non BPS D branes,
which are stable at $R>2$ and $R<2$ respectively. At $R=2$ there is a
perturbative enhancement of gauge symmetry and the two types of 
branes are related
by a global bulk symmetry transformation.  At this point in the moduli
space the associated  boundary states are
constructed  using a free fermion representation of the theory. Some
aspects of the enhancement of gauge symmetry in the S-dual type $\tilde I$
theory are discussed. 

%%%%%%%%%%%%%%%%%%%%%%%%%%%%%%%%%%%%%%%%%%%%%%%%%%%%%%%%%%%%%%%%%%%
\noblackbox
\baselineskip 20pt plus 2pt minus 2pt

\Date{July 2000}
\noblackbox

\newsec{Introduction}
It is well known that there are no points of enhanced gauge symmetry
in the moduli space of toroidally compactified type II superstring
theories.  Non-Abelian gauge symmetries for type II strings arise however in
  asymmetric
orbifolds \vafaa\dolan\   or
in free fermionic constructions \tyea\antona\ferraraa. Possibly the 
simplest examples of such  asymmetric orbifolds
are given by type II on  $T^d/(-1)^{F_L}\sigma_{1/2}$ where $F_L$ is the
left-moving space-time fermion 
number and $\sigma_{1/2}$ is a shift of order two in the
compactification lattice $\Gamma^{d,d}$.   
 $(-1)^{F_L}$ projects out the supercharges  coming from the
leftmovers and the freely acting  lattice shift $\sigma_{1/2}$ ensures
that no supersymmetries will be  
reintroduced in the twisted sector. Hence such a theory has only sixteen
supersymmetries  and has $(4,0)$ space-time supersymmetry (where the
numbers count 
right and leftmoving space-time supersymmetry charges  in four
dimensional units).  Unlike in the case of the heterotic string there
is  one  leftmoving worldsheet supersymmetry
which remains unbroken together with four rightmoving ones as is necessary
for $N=4$ space-time supersymmetry \banksa.

D-branes \pola\ are extended objects on which open strings can end. 
They impose boundary conditions on worldsheet fields relating the left
and rightmovers. In type II and type I theories D-branes can be BPS
objects, i.e. they preserve half of the space-time supersymmetries.
Since  the asymmetric orbifolds discussed above do not have any leftmoving
supersymmetries  D-branes can not be
BPS objects. Recently, there has been a great interest in nonsupersymmetric
string theory and non BPS branes (see for example
\sena\schwarza\gaberdiela\lerdar\ and 
references therein). In the following we will discuss some of the
properties of non-BPS D-branes in the simplest example of the  asymmetric
orbifold defined above, which is type IIA or IIB compactified on
$S^1/(-1)^{F_L}\sigma_{1/2}$, where $\sigma_{1/2}$ is the shift $X\to X+\pi
R$ and  $R$ is the radius of the circle.

Various aspects of D-branes in asymmetric orbifolds are  discussed in the
literature: D-branes in asymmetric orbifolds constructed using T-duality
\brunnera, using Sagnotti's method of open descendants \bianchic,
asymmetric orientifolds \sagnottik\carloa\ and  asymmetric type I theories \blumenhagenb.
\newsec{Closed string spectrum}
Firstly  consider the compactification of type IIA/B on a circle of radius
$R$. Identifying  $X^9\sim X^9+2\pi R$. The compactified momenta
$(p_L,p_R)$ lie on a lattice $\Gamma_{1,1}$ (setting $\alpha'=2$).
\eqn\latone{p_L= {m\over R}+{n R\over 2},\quad p_L= {m\over R}-{n R\over
2}.}
In this units the self-dual radius, i.e. the radius where the closed bosonic
string has enhanced $SU(2)\times SU(2)$ gauge symmetry, is
given by $R=\sqrt{2}$. Note that in a  toroidal compactification of type
II there are no points of enhanced non-Abelian gauge symmetry, because the GSO
projection removes  from the spectrum all the massless states carrying
only lattice momentum and winding 
which are necessary for symmetry enhancement. 
The one loop partition function of type IIA/B  is given by
\eqn\oneloopone{Z_{1,1}= \int {d^2\tau\over \tau_2} \tau_2^{-9/2}
{1\over |\eta(\tau)|^{16}}\big(\chi_{8,v}- 
\chi_{8,c}\big)\big(\bar\chi_{8,v}-\bar\chi_{8,{s/c}}\big)\sum_{p\in
\Gamma_{1,1}}q^{\half p_L^2}\bar q^{\half p_R^2}.}
Here $\chi_{8,o},\chi_{8,v},\chi_{8,s},\chi_{8,c}$ are  $SO(8)$
characters corresponding to adjoint,vector, and the two conjugate
spinor representations 
respectively.

String orbifolds are  constructed from a closed string theory by projecting
onto invariant states and adding twisted sectors \dixona. The main consistency
requirement of this procedure is modular invariance of the torus partition
function \vafab. 
At any radius $R$ the asymmetric $Z_2$ orbifold can be defined by
projecting with  $g=(-1)^{F_L}
\sigma_{1/2}$ where $F_L$ is the leftmoving space-time fermion number and
$\sigma_{1/2}$ is the shift by half the period of the circle $ X\to X+\pi R$.
Defining for  lattice vectors $l\in \Gamma_{1,1}$,  $l=(l_L,l_R)$ the
inner product is given by 
$l\cdot l^\prime = l_L\cdot l^\prime_L- l_R\cdot l^\prime_R$
The half period shift is then encoded in the lattice vector
$v=(R/4,-R/4$). 

For definiteness we will consider the IIB compactification here.
The twist in the time  direction in the path integral is
defined by 
\eqn\onelooponb{Z_{(g,1)}= \tr\big((-1)^{F_R} e^{2\pi i v\cdot p} q^{L_0}\bar
q^{\bar L_0}\big).}
Adding \oneloopone\ and \onelooponb\ gives the untwisted contribution to
the orbifold partition function
\eqn\untwis{\eqalign{Z_{untw}=&{1\over 2}(Z_{(1,1)}+Z_{(g,1)})\cr
=& \int {d^2\tau\over \tau_2} \tau_2^{-9/2} {1\over
|\eta(\tau)|^{16}}\big(\chi_{8,v}- 
\chi_{8,c}\big) \bar\chi_{8,v} \sum_{p\in \Gamma_{1,1}} \big(1+ e^{2\pi i
p\cdot v}\big) q^{\half p_L^2}\bar q^{\half p_R^2}\cr 
&- \int {d^2\tau\over \tau_2} \tau_2^{-9/2}{1\over
|\eta(\tau)|^{16}}\big(\chi_{8,v}- 
\chi_{8,c}\big) \bar\chi_{8,c} \sum_{p\in \Gamma_{1,1}} \big(1- e^{2\pi i
p\cdot v}\big) q^{\half p_L^2}\bar q^{\half p_R^2}
.}}
The twisted part of the partition function can be determined by modular
transformation since $Z_{(g,1)}(-1/\tau)=Z_{(1,g)}(\tau)$ and
$Z_{(1,g)}(\tau+1)=Z_{(g,g)}(\tau) $, the twisted sector contribution is
found to be
\eqn\twisc{\eqalign{Z_{tw}=& {1\over 2}(Z_{(1,g)}+Z_{(g,g)})\cr
=& \int {d^2\tau\over \tau_2} \tau_2^{-9/2}{1\over
|\eta(\tau)|^{16}}\big(\chi_{8,v}- 
\chi_{8,c}\big) \bar\chi_{8,o} \sum_{p\in \Gamma_{1,1}+v} \big(1- e^{2\pi i
p\cdot v}\big) q^{\half p_L^2}\bar q^{\half p_R^2}\cr
&- \int {d^2\tau\over \tau_2} \tau_2^{-9/2}{1\over
|\eta(\tau)|^{16}}\big(\chi_{8,v}- 
\chi_{8,c}\big) \bar\chi_{8,s} \sum_{p\in \Gamma_{1,1}+v} \big(1+ e^{2\pi i
p\cdot v}\big) q^{\half p_L^2}\bar q^{\half p_R^2}.}}
The complete partition function is then given by
$Z=Z_{untw}+Z_{tw}$. The momenta in the four different sectors are
given by 
\eqn\momlat{\eqalign{  \Lambda_1=\{p\in \Gamma_{1,1}, (-)^{p.v}=+1\} :& \; 
(p_L= {2m\over R}+n{R\over 2}, \; p_R= {2m\over R}-n{R\over 2}),\cr
 \Lambda_2=\{p\in \Gamma_{1,1}, (-)^{p.v}=-1\} :&\;  (
p_L= {2m+1\over R}+n{R\over 2}, \;p_R= {2m+1\over R}-n{R\over 2}),\cr
   \Lambda_3=\{p\in \Gamma_{1,1}+v, (-)^{p.v}=+1\}:&\;  
(p_L= {2m\over R}+(n+\half ){R\over 2},\; p_R= {2m\over
R}-(n+\half){R\over 2}),\cr 
    \Lambda_4=\{p\in \Gamma_{1,1}+v, (-)^{p.v}=-1\}:& \; (p_L=
{2m+1\over R}+(n+\half){R\over 2},\; p_R= {2m+1\over R}-(n+\half){R\over
2}).}}   
An important feature of this partition function is that at $R=2$ there are
additional massless states coming from the twisted sector \twisc\ in
$(\chi_{8,v}-\chi_{8,c})\bar\chi_{8,o}$. These 
states are responsible for the  enhanced $SU(2)$ gauge symmetry. The closed
string spectrum has $(4,0)$ space-time supersymmetry, which is evident from
the appearance of  $\chi_{8,v}-
\chi_{8,c}$ in the partition function $Z$.

The closed string spectrum can equivalently be described \vafaa\ by
replacing the 
standard leftmoving GSO projection $(-1)^{f_L}=+1$ with a modified
projection  $(-1)^{f_L+p\cdot 
v}=+1$, where $f_L$ is the leftmoving worldsheet fermion number.

\newsec{Non-BPS D-branes}
A very useful tool in the description of D branes is provieded by 
 boundary states
\polcai\callana\greena\greenb. There are several conditions which are
imposed on
boundary states in string theories.
The boundary state  has to preserve (super)-conformal
invariance. On the half plane this is equivalent to continuity conditions
on the stress tensor and its  superpartner, which translate into the following
conditions on the boundary states,
\eqn\boundcona{(L_n-\bar L_{-n})\mid B,\eta \rangle =0,\quad (G_k+i\eta \bar
G_{-k})\mid B,\eta\rangle =0,}
where $\eta=\pm 1$ is related to the spin structure for  worldsheet
fermions. 
In most cases the string background  has a simple  (factorized)
dependence on the ghosts and 
superghosts. Then (super) conformal invariance \boundcona\ is equivalent 
 to BRST invariance
\eqn\brstc{(Q_{BRST}+\bar{Q}_{BRST})\mid B\rangle=0.}
It is important that the boundary state itself is not a state in the closed
string Hilbert space (it has infinite norm), but when expanded in terms of
closed string modes only states which appear in the closed string partition
function \oneloopone\ appear. Note that the closed string states
appearing in the 
boundary state are 'off shell' because they do not satisfy the
$L_0+\bar{L}_0$ mass 
shell condition (although \boundcona\ implies that they satisfy the level
matching condition).

An important consistency conditions on boundary states are the Cardy
 constraints \cardy\ which demand that the cylinder partition function
 constructed 
\ from a boundary state  has an open string interpretation. This means that
under world sheet duality the cylinder partition function maps into an open
string partition function.

The boundary states will be constructed in a lightcone frame for
simplicity, the covariant construction of D-brane boundary states is
discussed (for example) in \devech.
In the NS-NS sector a boundary state (in the lightcone frame) will be of
the form
\eqn\boundsta{\eqalign{\mid B,\eta\rangle_{NSNS}&= {\cal
N}_{NSNS}\exp\Big(\sum_n {1\over n}\big(-a^i_{-n}\bar 
a^i_{-n}+a^a_{-n}\bar
a^a_{-n}\big)\Big)\cr
&\times \exp\Big(+i\eta \sum_r \big(-b^i_{-r}\bar
b^i_{-r}+b^a_{-r}\bar
b^a_{-r}\big) \Big)\mid B ,\eta\rangle_0         .}}
Here the indices $i$ and $a$ run over Neumann and Dirichlet directions
respectively. ${\cal N}_{NSNS}$ is a normalization factor which depends on
the boundary conditions in the compact and noncompact
directions. $\mid B\rangle_0$ denotes the zero mode part of the NS-NS 
boundary state. 
If the boundary condition in the compactified direction is
Neumann, the boundary state will contain a sum over winding modes. If it is
Dirichlet it will contain only a sum over even momentum modes. 
\eqn\dnndst{\eqalign{\mid N\rangle_0 &= \sum_n \mid p_L= {nR\over 2},
p_R=-{nR\over 
2}\rangle,\cr
 \mid D,x\rangle_0 &= \sum_m e^{i 2mx/R}\mid p_L= {2m\over R}, p_R={2m\over
R}\rangle.}}
The NS-NS boundary state has to be invariant under the right-moving GSO
projection $\half(1-(-1)^{f_R})$. This selects the following combination of the
$\eta=\pm1$ boundary states
\eqn\boundsb{\mid B\rangle_{NSNS}= \mid B, +\rangle_{NSNS}- \mid B,
-\rangle_{NSNS}.} 
Note that in the untwisted NSNS sector it is possible to impose either N or D
boundary conditions on the compact coordinate. Since all states appearing
in \dnndst\ have even $p\cdot v$ the boundary states also satisfy the modified
leftmoving GSO projection.

In the R-R sector a boundary state (in the lightcone frame) will be of
the form
\eqn\boundsta{\mid B,\eta\rangle_{RR}=  {\cal N}_{RR}\exp\Big(\sum_n
{1\over n}\big(-a^i_{-n}\bar 
a^i_{-n}+a^a_{-n}\bar
a^a_{-n}\big)+i\eta \sum_r \big(-d^i_{-n}\bar
d^i_{-n}+d^a_{-n}\bar
d^a_{-n}\big) \Big)\mid \eta \rangle^0_{RR}         .}
Defining 
\eqn\zmdef{d_{\pm}^\mu ={1\over \sqrt{2}}\big( d_0^\mu\pm i
\bar{d}_0^\mu\big). } 
Where the zero mode piece satisfies 
\eqn\zmpic{\eqalign{d_\eta^i\mid \eta \rangle^0_{RR}&=0,\quad i=Neumann,\cr
d_{-\eta}^a\mid \eta \rangle^0_{RR}&=0,\quad a=Dirichlet.}}
Since $(-1)^{f_R}\mid B,\eta\rangle = \mid B -\eta\rangle$, invariance
under the 
rightmoving GSO projection implies that the RR part of the boundary state
is of the following form 
\eqn\rrcomb{\mid B \rangle_{RR}= \mid B,+\rangle_{RR}+\mid B,-\rangle_{RR}.}
The RR potentials which appear, together with the boundary conditions
in the compact directions are determined by the closed string
partition function. Only states which appear in the closed string
partition function $Z$ (defined in \untwis\ and \twisc) can be used to
construct the boundary states. 
\medskip
In the partition function there are two sectors with RR-fields, one coming
\ from the untwisted sector \untwis, $\chi_{8,c}\bar\chi_{8,c}$ and one
coming from $\chi_{8,c}\bar\chi_{8,s}$ in \twisc.

Since the momenta associated with $\chi_{8,c}\bar\chi_{8,c}$ sector lie in
 $\Lambda_2$,  Neumann boundary
conditions in the compact directions $p_L=-p_R$ are inconsistent.
 Only  Dirichlet boundary 
conditions $p_L=p_R$ will be consistent.

The generalized leftmoving GSO projection $(-1)^{f_L+p\cdot v}=+1$ acting
on the boundary state selects RR field strengths with $p$ odd (in agreement
with the field content in $\chi_{8,c}\bar\chi_{8,c}$).
The compact momentum part of the boundary state is.
\eqn\Brama{\mid B\rangle_{0,RR}^D = \sum_m e^{i(2m+1)x/R} \mid p_L=
{2m+1\over R},p_R= 
{2m+1\over 
R}\rangle\otimes \mid p=2k-1\rangle_0 . }
The boundary state with Dirichlet boundary conditions in the compact
directions has a simple interpretation as a superposition of a
$D(2k-1)$ brane at 
$x$ and a $\bar{D}(2k-1)$ brane at $x+\pi R$, since the action of
$(-1)^{F_L}\sigma_{1/2}$ reverses the RR charge and translates the position of
the brane by $\pi R$. The massless RR part of the boundary state cancel
between the brane and the antibrane and only massive RR states 
nonvanishing momenta contribute.

The second sector of  RR fields lies  in  the twisted sector \twisc\
$\chi_{8,c}\bar\chi_{8,s}$. Note that because of 
the momenta which appear in this sector  $\Lambda_3$, Dirichlet  boundary
conditions $p_L=p_R$ are inconsistent.   Only Neumann  boundary
conditions $p_L=-p_R$ will be consistent. 
Since the momenta in $\Lambda_3$ have even $p\cdot v$, the generalized
leftmoving GSO projection selects RR field strengths with $p$ even
(in agreement 
with the field content in $\chi_{8,c}\bar\chi_{8,s}$).
 The compact momentum part of the boundary state is .
\eqn\Bramb{\mid B\rangle_{0,RR}^N = \sum_n \mid p_L=
{(n+\half)R/2},p_R=-(n+\half)R/2\rangle \otimes \mid p=2k\rangle_0. }
Again this boundary state can be interpreted as a superposition of a
$D(2k)$ and $\bar{D}(2k)$ wrapped on the circle with a nontrivial Wilson line.
\newsec{Open string spectrum}
The boundary states can be used to construct a cylinder diagram which has
the interpretation of a closed string tree level  exchange diagram. World
sheet duality turns the cylinder into an annulus, which has the
interpretation of a one loop open string diagram and hence determines
open string spectrum. 

 The contribution to the cylinder amplitude coming from the NS-NS
sector will depend on whether we have D or N boundary conditions on the
circle.
\eqn\znnn{\eqalign{Z^N_{NS}&= _{NS} \langle B,N\mid \int dl e^{-(\pi l
(L_0+\bar 
L_0-1))}\mid B,N\rangle_{NS} \cr
&=2\big({\cal N}_{NSNS}^N\big)^2 \int dl \;l^{(p-8)/2} {1\over
\eta^{12}(q)}\big(\theta_3^4(q)-\theta_4^3(q)) \sum_n e^{-\pi l {n^2 
R^2/4}}  }} 
\eqn\znnd{\eqalign{Z^D_{NS}&= _{NS} \langle B,D\mid \int dl e^{-(\pi l(L_0+\bar
L_0-1))}\mid B,D\rangle_{NS} \cr
&= 2\big({\cal N}_{NSNS}^D\big)^2\int dl \; l^{(p-8)/2}{1\over
\eta^{12}(q)}\big(\theta_3^4(q)-\theta_4^3(q)) \sum_m e^{-\pi l 
{4m^2/R^2}} . }}
Where $q=e^{-2\pi l}$ and   $p+1$ denotes the number of noncompact
Neumann directions.  The contribution for Neumann boundary condition
from the RR part is 
\eqn\zrrn{\eqalign{Z^N_{RR}&= _{RR} \langle B,D\mid \int e^{-(\pi t (L_0+\bar
L_0))}\mid B,D\rangle_{RR} \cr
&= {1\over 8}\big({\cal N}_{RR}^N\big)^2\int dl\; l^{(p-8)/2} {1\over
\eta^{12}(q)}\theta_2^4 (q)\sum_m e^{-\pi l 
{(m+1/2)^2R^2/4}} . }}
The contribution to the cylinder amplitude coming from the RR sector for
Dirichlet boundary condition on the circle are
given by
\eqn\zrrd{\eqalign{Z^D_{RR}&= _{RR} \langle B,D\mid \int dt e^{-(\pi l
(L_0+\bar 
L_0))}\mid B,D\rangle_{RR} \cr
&= {1\over 8}\big({\cal N}_{RR}^D\big)^2 \int dl\; l^{(p-8)/2} {1\over
\eta^{12}(q)}\theta_2^4(q) \sum_m e^{-\pi l 
{4(m+1/2)^2/R^2}}  }}
There are now two possible boundary states which are associated with having
either Neumann or Dirichlet boundary conditions in the compact direction.
\eqn\partfnd{Z^D= Z_{NS}^D+Z_R^D,\quad Z^N=Z_{NS}^N+Z_R^N.}
The open string partition functions are obtained by modular transformation
$ t = 1/l$. For the Dirichlet boundary conditions one finds
\eqn\zdfomc{\eqalign{\tilde Z^D&= R \int {dt\over
t}{t}^{-(p+1)/2}
{1\over \eta^{12}(w)}\Big( ({\cal
N}^D_{NSNS})^2(\theta_3^4(w)-\theta_2^4(w))\sum_m  
e^{-\pi t R^2m^2/4}\cr
&\quad\quad +{1\over 16}({\cal N}^D_{RR})^2\theta_4^4(w) \sum_m (-1)^me^{-\pi t
R^2m^2/4}\Big)\cr 
&= {V_{p+1}\over (4\pi)^{p+1}} \int {dt\over
t}{t}^{-(p+1)/2}{1\over \eta^{12}(w)}\Big( \chi_{8,v}\sum_m
e^{-\pi tR^2m^2}+\chi_{8,o}\sum_m
e^{-\pi t R^2(m+1/2)^2}\cr
&\quad\quad -\chi_{8,c} \sum_m e^{-\pi t
R^2m^2}-\chi_{8,s} \sum_m e^{-\pi t
R^2(2m+1)^2/4}\Big).}}
Where $w=e^{-2\pi t}$ and the normalization factors are determined by
demanding that \zdfomc\ has the interpretation of an open string partition
function $\tilde Z= \tr(w^{H}{\cal P})$ where the generalized GSO
projection is 
given by ${\cal P}=\half(1+(-1)^{f+m^2})$. Here odd winding numbers
correspond to strings winding around half the circle of radius R, i.e. open
strings
starting at a brane at $x=x_0$ and ending at an antibrane at
$x=x_0+\pi R$. The normalization 
factors are the  determined
\eqn\normla{({\cal N}^D_{NSNS})^2= R {V_{p+1}\over
(4\pi)^{p+1}},\quad ({\cal N}^N_{RR})^2= -{ 16}\;{R}\; {V_{p+1}\over
(4\pi)^{p+1}}.} 
For the Neumann boundary conditions one finds
\eqn\zdfomd{\eqalign{\tilde Z^N&= {4\over R}\int {dt\over
t}{t}^{-(p+1)/2}
{1\over \eta^{12}(w)}\Big( ({\cal
N}^N_{NSNS})^2(\theta_3^4(w)-\theta_2^4(w))\sum_m  
e^{-\pi t 4m^2/R^2}\cr
&\quad \quad +{1\over 16}({\cal N}^N_{RR})^2\theta_4^4(w) \sum_m
(-1)^me^{-\pi t 
4m^2/R^2}\Big)\cr
&= {V_{p+1}\over (4\pi)^{p+1}}\int {dt\over
t}{t}^{-(p+1)/2}{1\over \eta^{12}(w)}\Big( \chi_{8,v}\sum_m
e^{-\pi t 16m^2/R^2}+\chi_{8,o}\sum_m
e^{-\pi t  16(m+1/2)^2/R^2}\cr
&\quad \quad -\chi_{8,c} \sum_m e^{-\pi t
16m^2/R^2} -\chi_{8,s} \sum_m e^{-\pi t
4(2m+1)^2/R^2}\Big).}}
The  open string amplitude is  consistent with the following
generalized GSO projection ${\cal P}=\half(1-(-1)^{f+m^2})$. This determines
\eqn\normlb{({\cal N}^N_{NSNS})^2= {4\over R} {V_{p+1}\over
(4\pi)^{p+1}},\quad ({\cal N}^N_{RR})^2=- {64 \over R} {V_{p+1}\over
(4\pi)^{p+1}}.}

Note that in certain ranges of the radius $R$ there is a tachyon in the
spectrum coming from the $\chi_o$ sector. However there is no tachyon in
$\tilde Z_D$ for $R>2$ and there is no tachyon in $\tilde Z_N$ for
$R<2$. Hence there is 
always a non-BPS brane which is does not have a tachyon in the open string
spectrum and is therefore stable. Note also that the critical radius $R=2$
is the radius where an enhanced gauge symmetry in the bulk appears.
The two partition functions $\tilde Z^D$ \zdfomc\ and $\tilde Z^N$
\zdfomd\  are equal and they can be 
expressed as at that radius as
\eqn\rrtwo{Z\mid_{R=2}={V_{p+1}\over (4\pi)^{p+1}}\int {dt\over
t}{t}^{-(p+1)/2} {1\over \eta^{12}(w)}\Big( \theta_3^5(w)-\theta_4^5(w)
-\theta_2^4(w) \theta_3(w)\Big).}
The 
contribution to the partition function coming from the massless sector
vanishes, which indicates a degeneracy between the number of bosons and
fermions in  the massless spectrum.  However neither is there such a
degeneracy at higher levels nor are the massless interactions supersymmetric.

\newsec{Free fermion construction}
A compactification of type II strings from ten
to $d$ dimensions uses an internal SCFT with central charge $c= 3/2(10-d)$ 
The free fermionic construction \tyea\antona\   uses an  internal
conformal system is made of  $3(10-d)$ free left and
rightmoving fermions. Different models come from the choice of spin
structures (or equivalently generalized GSO) projections, consistent with
modular invariance, spin statistics and factorization. Since the left and
rightmoving fermions can have different generalized GSO projections many
interesting models with various  numbers of left and rightmoving
spactime supersymmetries and gauge groups  are possible to construct
\ferraraa\tyec.  

The simplest case is given by a compactification to nine dimension using
 three real fermions. Indeed  the asymmetric orbifold discussed in section
 4 at radius $R=2$  can be
 represented using  free fermions. The internal fermion $\psi_9$ and 
 the internal boson $X_9$ are replaced by three real fermions
$\lambda_1,\lambda_2,\lambda_3$, via $\psi_9=\lambda_1, \partial
X_9=\lambda_2\lambda_3$.  Hence the worldsheet fields in the nine
 dimensional theory 
are given by 
\eqn\wsfields{left:\psi^\mu,\lambda_1,\lambda_2,\lambda_3\quad \quad right:
 \bar\psi^\mu,\bar \lambda_1,\bar 
\lambda_2,\bar \lambda_3, \quad \quad \mu=0,\dots,8.}
The worldsheet supercurrents are given by
\eqn\supcurr{T_F= i\big(\sum_{\mu=0}^8\partial_z X^\mu \psi_\mu+
 \lambda_1\lambda_2\lambda_3\big), \quad \bar T_F= 
i\big(\sum_{\mu=0}^8\partial_{\bar z} X^\mu\bar \psi_\mu
 +\bar\lambda_1\bar\lambda_2\bar\lambda_3\big).} 
In the free fermionic construction of the asymmetric orbifold compactification
\antona\ one selects two sets of fermions 
$S=\{\psi^\mu,\lambda_1\}$ and $F=\{ \lambda_2,\lambda_3, \bar \lambda_1,
 \bar \lambda_2, \bar  \lambda_3, \bar\psi^\mu\}$.
There are four sectors $0,F,S, F/S$ and two GSO projections $(-1)^F=+1$ and
$(-1)^S=+1$. The closed string partition function is then  given by

\eqn\lattthetc{\eqalign{Z= & \int {d^2\tau\over
\tau_2^2}\tau_2^{p/2}{1\over |\eta(\tau)|^{16}}\big(\chi_{8,v}- 
\chi_{8,c}\big) {1\over \bar\eta^4}\Big( \bar \theta_3^4
\left|\theta_3\right|^2-\bar 
\theta_4^4\left|\theta_4\right|^2- \bar\theta_2^4
\left|\theta_2\right|^2\Big).}}
In the NS-NS sector the boundary conditions can be encoded as
\eqn\bndcff{\eqalign{(\psi^i_r+i\eta_\psi\bar \psi_{-r}^i)\mid
B\rangle&=0,\quad i=Neumann,\cr
(\psi^a_r-i\eta_\psi\bar \psi_{-r}^a)\mid
B\rangle&=0,\quad a=Dirichlet,\cr
  \quad(\lambda^i_r+i M^i_j\bar\lambda^j_{-r})\mid B\rangle&=0.}}
Where $\eta_\psi=\pm 1$ and $M$ is an $O(3)$ matrix. By a suitable choice
of basis $M$ can be expressed as $M= diag(\eta_1,\eta_2,\eta_3)$ with
$\eta_i=\pm 1$. These choices determine the 
boundary state $\mid B,\eta_\psi,\eta_1,\eta_2,\eta_3\rangle$. 
The condition of worldsheet superconformal invariance  \boundcona\  with
$T_F$ defined by \supcurr\ implies the following condition
\eqn\condeta{\det(M)= \eta_1\eta_2\eta_3=\eta_\psi.}
 The boundary state for the internal fermions $\lambda_i$ can be
represented as a coherent state imposing the conditions \bndcff.
The action of the right-moving GSO projection $(-1)^{S}$ acts as
$(-1)^{f_R}$ (where $f_R$ is the worldsheet fermion number associated
with $\psi_\mu,\lambda_1$) on such a boundary 
state 
\eqn\gsprjb{(-1)^{S}\mid B,\eta_\psi, \eta_1,\eta_2,
\eta_3\rangle_{NS} = - \mid B,-\eta_\psi, -\eta_1,\eta_2,
\eta_3\rangle_{NS}. } 
The other GSO projection in this model is $(-1)^F$ and involves both
$\lambda_2,\lambda_3$ and $\bar\lambda_2,\bar \lambda_3$ 
and does therefore not change the sign of $\eta_2,\eta_3$ and acts
effectively as $(-1)^{f_L}$. One  possible
boundary state in the NS-NS sector is given by\foot{The normalizations
of the boundary states 
are determined in the same way as in section 4, and given by  setting $R=2$.}
\eqn\bnsnsp{\mid B\rangle^N_{NSNS}= \mid B, 1,1,1,1\rangle_{NS}- \mid B,
-1,-1,1,1\rangle_{NS}.}
Note that this choice of signs is equivalent to imposing Neumann boundary
conditions on the original compact boson $X^9$. A different choice is 
\eqn\bnsnspb{\mid B\rangle^D_{NSNS}= \mid B, 1,-1,-1,1\rangle- \mid B,
-1,1,-1,1\rangle,}
which corresponds to Dirichlet boundary condition on the  boson $X^9$. 
Note that the two boundary conditions are related by an $SO(3)$ rotation of
the boundary conditions which acts on the matrix $M$ as 
\eqn\mmulpt{M\to \pmatrix{-1&&\cr &-1&\cr &&1}M,}
which send $\eta_{1,2}\to -\eta_{1,2}, \eta_3\to \eta_3$ and hence maps
\bnsnsp\ into \bnsnspb.
For both boundary states
 the NS-NS part of the cylinder partition function is given by
\eqn\cylpns{Z_{NS}= \langle B\mid \Delta \mid B\rangle_{NSNS}= \int
dl\; l^{(8-p)/2} {1\over 
\eta^{12}(q)}\big( \theta_3^4(q)-\theta_4^4(q)\big)\theta_3(q). } 
In the RR sector the boundary state can be buildt as a coherent state acting
on a zero mode part. The zero mode part is constructed by creation and
annihilation operators 
\eqn\creaanni{\eqalign{\psi_\eta^\mu&=\psi^\mu_0+i\eta \bar\psi^\mu_0,\cr 
\lambda^i_{\eta}&= \lambda^i_0+i\eta_i\bar\lambda^i_0, \quad i=1,2,3.}}
The zero mode piece will then be (in the lightcone gauge) a bispinor of
$SO(10)$. Hence these states will  be massive.
The zero mode part $\mid B, \eta_\psi,\eta_1,\eta_2\eta_3\rangle^{RR}$
will satisfy 
\eqn\szmrr{\eqalign{\psi^\mu_{\eta_\psi}\mid
B\rangle=0,\quad \mu=1,\cdots,7\cr
 \lambda^i_{\eta_i}\mid B\rangle=0,\quad i=1,2,3.}}
The GSO projection $(-1)^S$ implies that only the combination
\eqn\brrproj{\mid B^{RR}\rangle= \mid B,
\eta_\psi,\eta_1,\eta_2,\eta_3\rangle^{RR}+ \mid B,
-\eta_\psi,-\eta_1,\eta_2,\eta_3\rangle^{RR}}
is physical. The projection $(-1)^F$ determines which bispinors of
$SO(10)$  appear in \brrproj, depending on the choice of Dirichlet and
Neumann boundary conditions in the 9-th direction.

The RR contribution to the cylinder partition function (with the
correct normalization of the 
boundary states) will be of the form $\theta_2^5/\eta^{12}$ and the complete
cylinder partition function is  given by
\eqn\cylptf{Z_{cyl}= {V_{p+1}\over( 4\pi)^{p+1}}\int dl l^{(8-p)/2}
{1\over \eta^{12}(q)}\Big 
( \big(\theta_3^4(q)-\theta_4^4(q))\theta_3(q)-\theta_2^5(q)\Big).}
After a modular transformation to the open string channel this partition
function maps into  \rrtwo.  
Hence the free fermions provide an equivalent representation of the
enhanced symmetry point $R=2$ of the asymmetric orbifold discussed in
section 4.

\newsec{Gauge symmetry and stability}
In the asymmetric orbifold discussed above there is an enhancement of
gauge symmetry in the 
bulk at $R=2$. A vertex operator for the $SU(2)$ gauge bosons will be of the
form (in the zero ghost picture)
\eqn\gaugeboson{V(\zeta)= \zeta^a_\mu\int d^2z  (\partial X^\mu
+ik_\nu\psi^\nu\psi^\mu)\bar{J}^a
e^{ikX}.}
Where the currents $\bar J^a$ form  a $SU(2)_2$ current
algebra.  In the free fermionic construction they are given by $\bar J^a=
{i\over 2} \epsilon^{abc}\bar\lambda_b\bar\lambda_c$. 
On shell gauge invariance  manifests itself in  closed string amplitudes in
the following way: replacing the wave function  $\zeta^\mu_a$ by
$k_\mu \epsilon^a$ turns   vertex operator into  a total
derivative, which implies that the corresponding closed string 
 amplitudes $\langle
V(k_\mu\epsilon^a)V_1V_2\cdots V_n\rangle$ vanishe. In the
presence of a worldsheet boundary/D-brane, the total derivative can
pick up  a 
boundary term. In the case at hand 
 such a boundary term acts  like an infinitesimal
SU(2) rotation of the 
boundary state.
\eqn\gaugebosonb{V(k_\mu \epsilon^a)= \epsilon^a\int d^2z \partial(\bar{J}^a
e^{ikX})=\epsilon^a\oint\bar{J}^a
e^{ikX}. }
This means that a gauge transformation in the bulk induces a shift shift in
the scalars living on the brane. This is reminiscent of \greend, where it
was shown that in the case of the  bosonic string with enhanced gauge symmetry
$G\times G$ the massless open string scalars act like Goldstone bosons
and 
effectively break the bulk gauge symmetry to $(G\times G)/G$. 
 If the brane does not
fill space-time the Goldstone bosons are localized on the worldvolume of a
brane of lower dimensions.

Defining the zero mode of the $SU(2)$ current, $J_0^a={i\over 2}
\epsilon^{abc}\sum_n 
\bar\lambda^b_{-n}\lambda^c_{n}$, a finite gauge transformation in the bulk
is equivalent to a constant condensate on the boundary $\mid B
,\epsilon\rangle= \exp(i\epsilon_a J_0^a)\mid B\rangle$.
The constant condensate modifies the boundary conditions on the $\lambda_i$
\eqn\modbcf{\lambda^i_{r}\mid B, \epsilon\rangle = 
M^i_j  e^{i\epsilon_a J_0^a}\bar\lambda^j_{-r}e^{-i\epsilon_a J_0^a} \mid
B, \epsilon\rangle= M^i_j R(\epsilon)^j_k\bar\lambda^k_{-r} \mid
B, \epsilon\rangle.}
Hence a finite boundary condensate parameterized by $\epsilon^a$
  is equivalent to
modified boundary conditions given by $M^\prime= M R$ where $R$ is the
$SO(3)$ rotation matrix associated with the adjoint action of
$g=\exp(i\epsilon_aJ^a_0)$. Hence the rotation matrix  $R= \exp(\epsilon^a
N^a)$ where the antisymmetric matrix $(N^a)_{ij}=\epsilon_{aij}$. This
implies that the Neumann and Dirichlet 
boundary conditions \bnsnsp\ and \bnsnspb\ are related by a marginal
deformation just like in \senc. This marginal deformation of the boundary
state is equivalent to a gauge transformation of the bulk theory.  
\newsec{S-duality  and type ${\tilde I}$ Theory}
The nine dimensional theory which is obtained by the asymmetric orbifold
$(-)^{F_L}\sigma_{1/2}$ of type IIB has sixteen real supersymmetries. 
At generic values of the compactification radius the  gauge group is $U(1)^2$ 
 and comes  from the Kaluza Klein reduction of the graviton
and the antisymmetric tensor in ten dimensions.
This theory   belongs to the ${\cal M}_{1,1}$
component of the moduli space of consistent theories  with 16 supercharges
in 9 dimensions \schwarza\bianchih\bianchig\witteng. This class of theories
has  another  description as
 an orientifold which is  called type ${\tilde I}$ theory and can be
obtained from type IIB 
compactified on a circle  by gauging
the $Z_2$ symmetry $\Omega \sigma_{1/2}$ 
\polc\parka, where $\Omega$ denotes worldsheet parity tranformation. 
There is also a T-dual description of type ${\tilde I}$ theory (called
${\tilde IA}$) which has two orientifold 8-planes of opposite charge
sitting at the two endpoints of an interval. An analysis of BPS and non
BPS branes in this theory using K-theory was given in \bgh.

The asymmetric $(-1)^{F_L}\sigma_{1/2}$ orbifold and type ${\tilde I}$
are related by 
an $S$ duality \parka\bianchia\  as follows. Since the two
perturbative symmetries of type IIB 
$(-1)^{F_L}$ and $\Omega$ are mapped into each other via $(-1)^F_L=
S\Omega S^{-1}$ \send, the orbifolds 
involving half period shifts are dual because the adiabatic argument of
 Vafa and Witten \wittend\ applies.

Note that under an $S$ duality of the parent IIB theory the fundamental
string and NS 5-brane are mapped to the D string and D 5-brane
respectively. This is in accord with the fact that in type ${\tilde I}$
theory has the same BPS D branes a type I theory \bgh, whereas in the
$(-)^{F_L}\sigma_{1/2}$ all 1/2-BPS objects are perturbative string states.
Some aspects of these theories and their relation were discussed in
\bianchia\wittenf\ and  the relation of BPS saturated
amplitudes one loop amplitudes  was  discussed in \bianchib\gregoria.

An interesting question is how the  enhanced $SU(2)$ gauge symmetry
of the $(-)^{F_L}\sigma_{1/2}$
orbifold at $R=2$ manifests itself in the dual  type ${\tilde I}$ theory
(or in the T-dual type ${\tilde I}A$ theory).
The masses of the states which become massless at the critical radius
$R=2$ in the partition function \twisc\ are (in the string frame)
\eqn\massmass{m^2= \big({R \over 4 }- {1\over R}\big)^2.}
Note that these states fall into short (256 dimensional)
supermultiplets.
After an S-duality transformation this mass formula for the states in type
$\tilde I$ has the form (in
the dual string frame)
\eqn\massmassb{\tilde m^2= \big(e^{-\tilde \phi} {\tilde R \over 4} -
{1\over\tilde R}\big)^2.} 
This indicates that the enhancement of gauge symmetry in the type
$\tilde I$ theory is caused by a D string becoming massless. For this
interpretation to be valid one has to continue 
\massmass\ to strong coupling, which is justified for BPS states.

After a T-duality on the circle the corresponding mass formula for the type
 $\tilde I A$ theory is given by
\eqn\massmassc{\tilde m^2= \big(e^{-\tilde \phi} {1 \over 4} -
 \tilde R\big)^2.}
Just like in the discussion of enhanced gauge symmetry in type $I^\prime$
 theory \wittenf\ the enhancement of the gauge symmetry should be visible
 small coupling and large radius in the type
 $\tilde I A$ theory, where a perturbative analysis should suffice.  Type
 $\tilde IA$ theory is an orientifold with a $O^+_8$ and   $O^-_8$
 orientifold plane at the end of an interval. Since the RR charges are
 canceled between the orientifold planes, no D8 branes have to be
 introduced (in contrast to type $I^\prime$).  Although the properties of
 $O^-_8$ orientifold planes in supergravity are not well understood, it is
 quite likely that their properties at least 
what the supergravity is concerned
 (i.e. not taking open strings into account) are well approximated by 16
 D8-branes on top of an $O^+_8$. Following \wittenf\ the string coupling
 will diverge near the orientifold planes making the $SU(2)$ gauge
 enhancement an intrinsically nonperturbative phenomenon in $\tilde IA$ theory.
It would be very interesting to see
whether this enhancement of gauge symmetry can be analyzed in the
$\tilde IA$ theory using the methods developed in \bagrsch, where presumably it comes from bound states of D0
branes near an orientifold plane.

A M-theory realization of these nine dimensional theories is given by a
compactification of M-theory on a Klein bottle, which is not purely
geometrical since the Klein bottle is realized as a freely acting orbifold
of the torus, where the geometric action is accompanied with a sign reversal
of the three form potential $C\to -C$ \parka.

\newsec{Discussion}

In this paper we have discussed the construction of non BPS D-branes in a
very simply nine dimensional symmetric orbifold, generated by
$(-1)^{F_L}\sigma_{1/2}$. The interesting features of these D-branes are
that although they do not carry any RR charge, the boundary state contains
a (massive) RR part which is important for consistency and stability. 
There are two kinds of D branes, one with Dirichlet boundary conditions in
the compact direction which is stable for $R>2$ and one with Neumann
boundary conditions which is stable for $R<2$.
At  the critical point $R=2$, when these D-branes become
unstable, there is an enhanced gauge symmetry in the bulk. The deformation
of the boundary conditions from Neumann to Dirichlet can then  be
interpreted as 
a global gauge transformation in the bulk.
Since the D-branes are non BPS they cannot simply be continued to strong
coupling. 

Since they are stable,  Euclidean D-branes will however
correspond to D-instantons, which can give dominant nonperturbative
contributions to certain processes. Such non-BPS D-instantons will
have sixteen 
fermionic zero modes and possibly contribute to $R^4$ terms. It would be
interesting to investigate the effects of these non-BPS instantons  further.

The asymmetric orbifold which was considered in this paper is a very 'mild'
one since it only involves a non geometric action $(-1)^{F_L}$. It would be
very interesting to try to generalize this to more complicated asymmetric
orbifolds. It might be that the fact that some non geometric asymmetric
orbifolds have free fermionic realizations at special points in their
moduli space will be very useful, since the boundary states are very easy
to construct for the free fermions and the consistency condition, i.e. the
imposition of the generalized GSO projections in addition to the Cardy
constraints could be  more managable.  
\medskip
{\bf Acknowledgements}
\medskip
I wish to thank to Costas Bachas for useful conversations and
encouragement. I  gratefully acknowledge the kind hospitality of the Korean
Institute for Advanced Study (KIAS) 
where part of this work was performed. This work is 
supported in part by the David and Lucile Packard
Foundation.

\listrefs

\end